\documentstyle[aps]{revtex}

\begin{document}

\title{  
     Instantons and radial excitations in attractive Bose-Einstein condensates 
      }

\author{Janusz Skalski }

\address{
 So\l tan Institute for Nuclear Studies,\\
ul. Ho\.za 69, PL- 00 681, Warsaw, Poland \\
 e-mail: jskalski@fuw.edu.pl, tel/FAX: (48 22) 621 60 85
 }

\date{27 October 2001}

\maketitle

\begin{abstract}
 Imaginary- and real-time versions of an equation for the condensate density 
 are presented which describe dynamics and decay of any spherical Bose-Einstein 
 condensate (BEC) within the mean field approach. 
 We obtain quantized energies of collective, finite amplitude radial 
 oscillations and exact numerical instanton solutions which describe quantum 
 tunneling from both the metastable and radially excited states of the 
 BEC of $^7$Li atoms.   
 The mass parameter for the radial motion is found different 
 from the gaussian value assumed hitherto, but the effect of this difference 
 on decay exponents is small. 
  The collective breathing states form slightly compressed harmonic spectrum, n=4 state lying lower 
  than the second Bogoliubov (small amplitude) mode. The decay of these states, if excited, 
  may simulate a shorter than true lifetime of the metastable state. 
 By scaling arguments, results extend to other attractive BECs.

\end{abstract}

{PACS number(s): 03.75.Fi, 05.30.Jp, 05.45.-a, 47.20.Ky }

\section{Introduction }

\noindent 
In a series of experiments \cite{Hul0,Hul1,Hul2,Hul3}, a nonuniform 
Bose-Einstein condensate (BEC) of $^7$Li atoms was formed and proved metastable for atom  
numbers $N < N_c$ with $N_c \approx 1300$, in agreement with theoretical predictions 
\cite{Rup,7Li,Dok}. Due to 
attraction between atoms, such condensate is bound to collapse when $N > N_c$, 
but it may collapse also for $N < N_c$  via quantum or thermal tunneling. 
Within the mean-field appraoch, BEC is described by one wave function. 
Our aim in this work was to find the exact mean-field description of the quantum tunneling 
in this simplest conceivable many-body system. 
 Our additional motivation is the recently demonstarted ability to control the interaction 
 of $^{85}$Rb atoms in BEC \cite{tune}, which opens a new perspective to systematic experimental 
 checks on quantum tunneling.  

Up to now, studies of the quantum tunneling of BEC 
 relied on assuming gaussian wave functions \cite{7Li}, at least in 
assigning the mass parameter \cite{Dok}. Strictly speaking, once the mean field equation 
is specified, there is no place for such an assumption: This equation, taking a form of 
the nonlinear field equation, by itself determines the dynamics. To find solutions 
which correspond to quantum decay we use the method of instantons, i.e. 
fields evolving in imaginary time \cite{Col}, carried over
to mean-field theories of many-body systems \cite{LNP,JN}. It gives 
the decay rate $\Gamma= A e^{-S}$, with the exponent $S$ being the action for the optimal 
mean-field instanton, called bounce. 
We find the exact instantons for spherical BEC by first transforming the original 
imaginary-time mean-field equation to an equation for the condensate density, 
and then solving it numerically. Having done that wa can check the gaussian 
 ansatz. 
 
 The real-time version of the obtained equation 
encompasses finite amplitude collective radial oscillations of BEC.  
Applying quantization rule we find 
 energies of radial eigenmodes. By using imaginary-time dynamics we also find periodic instantons 
  determining decay exponents of these breathing modes. 
 In this way, the collective dynamics of an attractive spherical condensate close to instability 
 is obtained 
  from the mean field equation.  
 
 We assume that the dynamics of BEC is governed by the 
 time-dependent Gross-Pitaevskii equation (GPE) \cite{GP}
\begin{equation}
\label{eqGP}
i \hbar \partial_t \psi = - \frac{\hbar^2}{2m} \nabla^2 \psi + 
 ( V_{trap} + g \mid \psi \mid ^2 ) \psi   ,
\end{equation}
where $V_{trap}$ is the static trap potential and $g=4\pi\hbar^2a/m$,
 with $a$ the s-wave scattering length and $m$ the atomic mass. The wave 
 function is normalized as 
 $\int d^3r \mid\psi({\bf r},t)\mid^2=N$, 
 with $N$ the total number of atoms in the condensate.

 In the present work we consider a harmonic, spherically symmetric trap 
 $V_{trap}=\frac{1}{2}m\omega_0^2 r^2$. This suggests choosing 
 $d_0=\sqrt{\hbar/m\omega_0}$ 
 as the unit of length, $1/\omega_0$ as the unit of time and $\hbar\omega_0$ as
 the unit of energy. We also change the normalization of the wave function 
 to unity, $4\pi\int\mid\psi(r,t)\mid^2r^2dr=1$. To simplify equations we 
 work with the function $\phi(r,t)=r\psi(r,t)$, for which the GPE reads 
\begin{equation}
\label{GP1}
 i\frac{\partial\phi}{\partial t} = - \frac{1}{2} \frac{\partial^2\phi}{\partial r^2} + 
 (\frac{1}{2}r^2 + K \frac{\rho}{r^2})\phi ,
\end{equation}
 where $K=4\pi Na/d_0$, and the generalised density $\rho(r,t)= r^2 \mid \psi(r,t)\mid^2 $. 
 For a stationary state, $\phi(r,t)=\phi(r)exp(-i\epsilon t)$,  
 $\epsilon$ being the single-particle energy or chemical potential. 
 For each $N<N_c$ there are two stationary states,  
 one metastable and another unstable, at the top of the energy 
 barrier \cite{Dok} (cf Fig. 1).

\section{Quantum tunneling and equations for condensate density }
 
\noindent 
Quantum tunneling of BEC is described by a specific solution 
 to the equation 
\begin{equation}
\label{eqsp}
 \frac{\partial\phi}{\partial \tau} - \frac{1}{2} \frac{\partial^2\phi}{\partial r^2} + 
 (\frac{1}{2}r^2+K \frac{\rho}{r^2})\phi =  \epsilon \phi 
\end{equation}
 obtained from (\ref{GP1}) by a transition to imaginary time, 
 $t\rightarrow -i\tau$ \cite{LNP,super}. 
 Now, the density $\rho(r,\tau)= \phi(r,-\tau)^* \phi(r,\tau) $ 
 as $\phi(r,t)^*$ is replaced by $\phi(r,-\tau)^*$ upon $t\rightarrow -i\tau$.
  This makes Eq.(\ref{eqsp}) nonlocal in time. 
 Bounce has to satisfy the boundary conditions of 1) periodicity, 
  $\phi(r,\tau_p/2)=\phi(r,-\tau_p/2)=\phi_0(r)=r\psi_0(r)$, with $\psi_0(r)$  
 the amplitude of the metastable state, and  
 2) barrier penetration, i.e. $\phi(r,\tau=0)$ has to be some state of BEC 
  at the other side of the potential barrier. 
Eq. (\ref{eqsp}) preserves both the normalization  
  $4\pi\int dr \rho(r,\tau)=1$ and the energy 
   \begin{equation}
   \label{mhf}
    {\cal E}=4\pi\int_0^{\infty}dr\{
 \frac{1}{2}\frac{\partial \phi(-\tau)^*}{\partial r}\frac{\partial \phi(\tau)}{\partial r}+
  \frac{1}{2}\rho r^2+\frac{K}{2}\frac{\rho^2}{r^2}\} ,
 \end{equation}
with $E=N{\cal E}$ the energy of the metastable state. For a bounce starting from the metastable state
  the period $\tau_p$ extends to infinity \cite{LNP,super,JN}. The decay exponent reads \cite{LNP}
\begin{equation}
\label{S}
 S= 4\pi N \int_{-\tau_p/2}^{\tau_p/2} d\tau \int dr \phi(-\tau)^*\frac{\partial \phi}
{\partial \tau} (\tau).
\end{equation}
Since Eq.(\ref{eqsp}) is real and the boundary value $\phi_0(r)$ may be taken real,  
 we assume real $\phi(r,\tau)$ in the following. 

 Now the point is to transform the nonlocal in time instanton equation (\ref{eqsp}) into an  
 evolution equation for the condensate density. A transformation of the real-time GPE (\ref{GP1}) 
 to a fluid-dynamic form provides an analogy, but is conceptually simpler. 

  The bounce equation (\ref{eqsp}), with the boundary conditions specified above, 
  splits into two time-local equations for the time-even
  density $\rho$ and the time-odd current 
  $j(r,\tau)= -1/2(\phi(-\tau)\partial_r\phi(\tau)-\phi(\tau)\partial_r\phi(-\tau))$
\begin{equation}
  \label{roj1}
\frac{\partial \rho}{\partial \tau} + \frac{\partial j}{\partial r} = 0 , 
\end{equation}  
\begin{equation}
  \label{roj2}
\frac{\partial j}{\partial \tau}+\frac{1}{4}\frac{\partial^3 \rho}{\partial r^3} - 
 \frac{\partial \Theta}{\partial r} -\rho[r+K\frac{\partial(\frac{\rho}{r^2})}{\partial r}]
 =0 ,
\end{equation}
where the kinetic energy density $\Theta=\partial_r\phi(-\tau)\partial_r\phi(\tau)=
 [1/4(\partial_r \rho)^2 - j^2]/\rho$.
 When $\rho$ is non-negative (which is very plausible in the present case, but not granted in general as 
 $\rho(r,\tau)=\phi(r,\tau)\phi(r,-\tau)$ not $\phi(r,\tau)^2$), one can define a regular, time-odd function 
 $\chi=-\frac{1}{2}(\ln \phi(\tau)-\ln \phi(-\tau))$ which allows decomposition 
$\phi=\sqrt{\rho}e^{-\chi}$.  
From this, the fluid-dynamic representation of (\ref{eqsp}) follows, with 
the velocity field $\partial \chi/\partial r = j/\rho$. 
However, even for arbitrary $\rho$, one can eliminate $j$ from Eqs.(\ref{roj1},\ref{roj2}), 
which is more convenient. 
Introducing $f(r,\tau)=\int_0^r \rho(r',\tau)dr'$, so that $j=-\partial f/\partial \tau$ 
 and $\rho=\partial f/\partial r$, we automatically fulfil (\ref{roj1}), and 
(\ref{roj2}) transforms to the equation for the primitive of the bounce density, 
 basic for the imaginary-time dynamics of spherical BEC-s: 
\begin{equation}
\label{bou}
\frac{\partial^2 f}{\partial \tau^2} - \frac{1}{4}\frac{\partial^4 f}{\partial r^4}
 + \frac{\partial}{\partial r}(\frac{\frac{1}{4}(\frac{\partial^2f}{\partial r^2})^2
 -(\frac{\partial f}{\partial \tau})^2}{\frac{\partial f}{\partial r}})
 +\frac{\partial f}{\partial r}[r+K\frac{\partial}{\partial r}
 (\frac{\frac{\partial f}{\partial r}}{r^2})]
 =0  . 
\end{equation}

 Notice, that the finite amplitude oscillations of BEC around the metastable minimum 
 are described by the real-time version of Eq.(\ref{bou}), in which  
 $\partial_{\tau}^2 f$ and $(\partial_{\tau} f)^2$ are replaced by $-\partial_t^2 f$ and 
 $-(\partial_t f)^2$, respectively. 

An alternative, global approach to quantum tunneling 
 is to look for a minimum of action (\ref{S})  
under the condition of constant energy (\ref{mhf}) and norm. Indeed, for a regular $\chi$, 
i.e. positive $\rho$, $S=
4\pi N \int_{-\tau_p/2}^{\tau_p/2}d\tau\int_0^{\infty} dr j^2/\rho $. 
Using Eq.(\ref{mhf}) and 
 introducing an observable $Q$ uniquely labelling states  
 along the barrier, explicitly $Q(\tau)=\langle r^2\rangle /N=4\pi\int_0^{\infty} dr \rho r^2$, we obtain the 
 following functional 
\begin{equation}
\label{func}
S[f]=2 N\int_{Q(0)}^{Q(\tau_p/2)} dQ \sqrt{2 B(Q) (V(Q)-{\cal E})}
\end{equation}
  which is minimized by the primitive of the bounce density (note, that $Q(0)<Q(\tau_p/2)$ for BEC). 
  The potential energy $V(Q)=V[\rho(Q)]$,  
\begin{equation}
\label{V}
V[\rho]=4\pi\int_0^{\infty} dr[\frac{(\partial_r\rho)^2}{8\rho}+\frac{1}{2}\rho r^2+\frac{K\rho^2}{2 r^2}] 
\end{equation}
and the effective mass parameter $B(Q)=B[f(Q)]$  
\begin{equation}
B(Q)=4\pi\int_0^{\infty} dr \frac{(\frac{\partial f} {\partial Q})^2}{\rho}
\end{equation}
are both the functionals of $f$.  Note that Eq.(\ref{func}) is invariant with respect to 
 a change of the controlling variable, as for any other such variable $q$, $B(q)=B(Q)(dQ/dq)^2$. 
The energy conservation (\ref{mhf}) implies that for bounce 
${\dot Q}^2=2(V(Q)-{\cal E})/B(Q)$,
 with ${\dot Q} = dQ/d\tau$, and therefore ${\ddot Q}=\frac{\partial}{\partial Q} [(V(Q)-{\cal E})/B(Q)]$. 

\section{Results and discussion  }

\noindent 
We have solved Eq.(\ref{bou}) using the variable $Q$ rather than $\tau$. An initial sequence of densities 
$\rho_s(r,Q_i)$, with 30-50 $Q_i$ points covering the barrier region, was constructed by minimizing $V[\rho]$ 
(\ref{V}) under the constraint $Q=Q_i$. These constrained stationary densities 
 were then improved upon iteratively. 
The details of $\rho^{1/2}/r$, the counterpart of $\psi$ of Eq. (\ref{eqGP}), are obtained more precisely when   
the $r^2$ behaviour near $r=0$ and the harmonic oscillator asymptotics at infinity are factored out
in Eq.(\ref{bou}). 
 In numerical work, we use a function $\alpha(r,\tau)$ such that $\rho(r,\tau) = r^2 e^{-r^2} e^{2\alpha(r,\tau)}$, and 
 properly express the term $\partial_{\tau}^2 f - \partial_r((\partial_{\tau} f)^2/\rho)$ - see {\it Appendix}. 
 
 We also performed the minimization of the functional (\ref{func}) 
 treating $\rho(r_j,Q_i)$ as independent variables. It turns out that 
 $\rho(r,\tau)$ thus obtained do not fulfil Eq.(\ref{bou}) accurately, but the accuracy in $S$  
 is better than 0.1\%. 

The numerical results have been obtained with physical data on $^7$Li  
adopted after 
the most accurate treatment up to date \cite{Dok}. 
These values give $K=- 5.74\times 10^{-3}\times N$, and we obtain the critical value $K_c$ between --7.2249 and --7.2255 (
 $N_c$ between 1258.7 and 1258.8). 

 The potential energy $V(Q)$ from the bounce solutions (Fig.1), nearly identical with $V[\rho_s(Q)]$ for 
 constrained stationary $\rho_s(r,Q_i)$, is very flat between $Q(0)$ and $Q(\tau_p/2)$ 
 for $N\approx N_c$. For smaller $N$, it becomes quite peaked
around the summit, and its fall on the side of small $Q$ becomes 
 very steep. All the obtained bounce solutions result from the small adjustment of the initial densities.     
 For larger $N_c-N$ (and increasing ${\dot Q}^2$ and ${\ddot Q}$ terms, cf Eq.(\ref{Edyn})) this adjustment 
 becomes gradually more difficult. 
 We could not obtain the exact solution of Eq.(\ref{bou}) (or even a constrained stationary 
 $\rho_s(r,Q_i)$ for small $Q_i$) for $N\leq 1200$. 

 The mass parameters $B(Q)$ from various instanton solutions (Fig.2)  
 are nearly identical which shows that there exists a universal inertia for  
 the radial collective motion of BEC close to instability. 
 This may be understood as a consequence 
  of a nearly static character of solutions $f(r,Q)$ in the limited range of $N(K)$ values of interest: 
  $f$ depends weakly on $N$ close to critical $N_c$. 

For a gaussian density with a variable width $b(\tau)$, 
$\rho(r,\tau)=\pi^{-3/2}b^{-3} r^2 exp(-r^2/b^2)$, 
 the current $j= {\dot b} (r/b)\rho$ (Eq.(\ref{roj1})), and the mass parameter 
 $B(b)=4\pi/b^2\int_0^{\infty} r^2 \rho dr = 
 3/2$ (cf \cite{7Li,BEmass}). Hence, as for the gaussian density $Q=3 b^2/2$, 
 one has $4QB(Q)=B(\sqrt{Q})=1$, as used in Ref.\cite{Dok}. 
 As seen in Fig.2, this is a 
 fair assumption near the metastable minima, much worse though for smaller 
 $Q$ around the barrier's summit. 
 The error in $S$ due to the gaussian value of $B$ amounts 
 to 3-4\% in the cases studied.

 The bounce "amplitude", $\sqrt{\rho}/r$ (Fig.3), differs from the constrained 
 stationary values $\sqrt{\rho_s}/r$ mostly near $r=0$ and for small $Q$, by up to 1\%. 
 Decay exponents calculated with the initial densities $\rho_s(r,Q_i)$ 
 are only up to 0.3\% larger than the exact ones. 
  
 Next we turn to collective radial oscillations of BEC and their decay rates. 
 Quantum tunneling from the excited state with energy $\hbar \omega_n$ 
 may be treated by solving Eq.(\ref{bou}) as before, except that now 
 the period $\tau_p$ must be finite and $N{\cal E}$ replaced by 
 $N{\cal E}_n=N{\cal E}+ \hbar\omega_n$. One has to fix the 
boundary values $\rho(r,\tau_p/2)$ and the energies $\hbar \omega_n$.  

 The energies $\hbar\omega_n$ of the excited quantum states follow 
 from the quantization condition  $S= 2 n \pi $ \cite{LNP2}, where  
 $S=2N\int_{Q(0)}^{Q(t_p/2)}dQ\sqrt{2 B(Q) ({\cal E}_n-V(Q))}$, cf Eq.(\ref{func}), 
 with the period $t_p$ depending on $\hbar\omega_n$. Essentially,  
the imaginary-time boundary value $\rho(r,\tau_p/2)$ should match 
 the real-time oscillatory solution $\rho(r,t_p/2)$ at the turning point $Q$ at which 
$V(Q)={\cal E}_n$. 

We could not solve the real-time version of Eq.(\ref{bou}) as accurately as for instantons. 
However, since the action $S$ is insensitive to the details of solution, the lowest 
 quantized $\hbar\omega_n$ are quite accurate. Their comparison to  
 frequencies of the small amplitude oscillations, obtained from the GPE 
 linearized around $\psi_0$ (so called Bogolyubov spectrum, see \cite{Bog}), 
 is interesting. 
 The first radial excitation, $\hbar\omega_1$, is nearly equal, but 
 little smaller than, the lowest   
 Bogolyubov mode corresponding to the small amplitude limit, 
 except that it may not exist for a too shallow $V(Q)$, like for $N=1255$. 
  The higher modes, $\hbar\omega_n$, $n=2,3,...$, come 
 out roughly at the multiples of $\hbar\omega_1$ (Table 1), 
 and thus represent nearly harmonic spectrum of collective radial 
 oscillations. Its $n=2,3,4$ states are lower 
 than the second Bogolyubov mode, lying slightly above $4\hbar\omega_0$. 

Using the quantized oscillation energies we have found periodic instantons and calculated 
decay exponents for some low collective radial excitations. Such instanton with a period $\tau_p$  
 is quite similar to the so-called "thermon" which describes thermal decay from the 
 metastable ground state at the temperature $kT=\hbar/\tau_p$ \cite{ChT}.
 This similarity is easily understood: Thermal decay goes via thermal excitation of 
  quantum states above the metastable ground state and successive tunneling from them through 
   the smaller barrier. At higher temperatures, details of the excitation spectrum and 
    tunneling become irrelevant and the thermal tunneling rate $\exp(-(E_b-E)/(\hbar\omega_0))$, 
    with $E_b=N{\cal E}_b$ the barrier energy, 
     replaces a thermal mixture of rates $\exp(-S_n)$ from a few lowest excited states 
     at small temperature. The critical temperature $T_c$ at which this happens is 
    usually determined as the one at which $\Gamma_{thermal}(T_c) = \Gamma_{thermon}(T_c)$.

 Numerical results are collected in Table 1. Those in columns 2-4 depend only on 
 $K$ and relate to all BEC-s with attractive interaction. 
 Decay rates $\Gamma$ from the metastable $^7$Li BEC were calculated for 
 $\omega_0=908.41$ s$^{-1}$ \cite{Hul1},   
 assuming prefactor $(\Gamma/\omega_0)e^{S}=(\omega_1/\omega_0) (15 S/2 \pi)^{1/2}$ \cite{7Li}, 
 with $\omega_1$ from the small amplitude limit. 
 The exact prefactor is difficult to calculate, but it must be of the same order as in Table 1. 
Obtained values of $\Gamma$ are very close to those of Ref.\cite{Dok}, where the effect of too small    
  $B$-s is fortuitously reduced by half by the interpolation overestimating $V(Q)$. 

 The decrease of the decay exponents $S_n$ with the mode number $n$  
 and their comparison to exponents $S$ for the ground state are seen in Table 1, col. 5. For example, 
  quantum decay of the $n=2$ radial mode in BEC with $N=1245$ is nearly as quick as for the 
   ground state of the $N=1255$ BEC.  
The crossover temperature $T_c$ at which the thermal decay begins to 
 dominate over the quantum tunneling is approximated, neglecting prefactors, as 
 $kT_c/(\hbar\omega_0)=(E_b-E)/S$, with $E_b=N{\cal E}_b$   
 the barrier energy, $E=N{\cal E}$ for the metastable state, and $E=N{\cal E}_n$ for the excited state. 
 For the latter, the meaning of critical temperature is extended in analogy with that for the ground state: 
 Suppose the condensate in the oscillation state $n$ and ask at which temperature $T_c(n)$ the thermal rate exponent 
  equals the instanton action $S_n$. 
 These ratios (Table 1, column 3) show that for presently measurable $\Gamma$ and 
 $\omega_0=908.41$ s$^{-1}$, corresponding to $T=6.94$ nK, 
 $T_c\approx 1$nK, both for ground state and collective radial oscillations. 
 We notice, that for all finite-$\tau_p$ instantons describing decay out of the $n-$th collective radial 
 excitation, $\hbar/\tau_p < T_c(n)$ so, 
 at their own thermon "temperature", they dominate over the thermal decay. 

 The quantity $S/N$ (Table 1, column 4) shows $(N_c-N)^{\xi}$ behaviour 
 close to $N_c$, with $\xi=5/4$ \cite{7Li,Dok}. For larger $N_c-N$ the 
 effective exponent slightly decreases to $\xi\approx 1.2$. 
 Consider now two different attractive BEC-s with critical particle numbers 
 $N_c$ and $N_c'$. The same decay exponent $S$ is obtained for 
 such (not too large) $N_c-N$ and $N_c'-N'$ which satisfy the relation 
 $N_c'-N'=(N_c-N)(N_c'/N_c)^{1-1/\xi}$, with $1-1/\xi\approx 1/5$. 
 Thus, e.g. we obtain $S=9.44$ for $N_c'-N'\approx 12$ in BEC with $N_c'=6294$, and 
 for $N_c'-N'=6-7$ in BEC with $N_c'=251.76$, cf Table 1. 

   In summary, the equations for the condensate density, describing both 
   the real- and imaginary-time 
   dynamics of spherical BEC, were formulated and the exact instanton solutions were found numerically, 
   also for collective radial excitations. The determined mass parameter (Fig. 2) deviates from 
 the gaussian ansatz, but the calculated decay exponents for the metastable states agree well with Ref.\cite{Dok}. 
 It follows from Fig.2 that the exact mass parametr may be more relevant to the behind-the-barrier collapse phase, 
i.e for smaller $Q$. 
 The quantized energies of collective finite amplitude radial vibrations form nearly harmonic 
  (slightly compressed) spectrum with $\omega_n\approx n \omega_1$, where $\omega_1$ is slightly lower than 
  the lowest Bogolyubov mode. 
  The $n=2,3,4$ collective oscillation states lie lower than the second Bogolyubov mode.  
  Any excitation (thermal or otherwise, e.g. by modulation of the trapping oscillator frequancy) of these 
  collective vibrations must lead to 
  a faster decay of the condensate, as the Table 1 shows. 

 If quantum tunneling is not to be overshadowed by thermal decay, the experiments should 
 proceed at low $T$ and/or large $\omega_0$. Since theoretical results are more certain 
 away from $N_c$, where the exponent $S$ dominates decay, one should probe a range of 
 moderate $S$, giving observable, but not too large $\Gamma$ (perhaps, by discarding prompt collapses). 
 The corresponding range of $N_c-N$ values depends on $N_c$ as $N_c^{1/5}$. 


\appendix

\section*{Numerical methods}

\noindent 
 We have $\rho(s,\tau) = s e^{-s} e^{2\alpha(s,\tau)}$ with $s=r^2$. 
The stationary GPE leads to the equation for $\alpha(s)$
\begin{equation}
\label{Estat}
2s (\frac{d^2\alpha}{ds^2}+(\frac{d\alpha}{ds})^2-\frac{d\alpha}{ds}) + 
 3\frac{d\alpha}{ds} - K e^{2\alpha-s} + \beta = 0 , 
\end{equation} 
with $\beta=\epsilon-3/2$.
For large $s$, $\alpha \approx \frac{\beta}{2}\ln s$
\cite{EB}. The normalization of $\phi$ implies  
 the next two terms   
\begin{equation}
\label{Eas}
 \alpha(s)=\frac{\beta}{2}\ln s+C-\frac{\beta
(\beta-1)}{4s}+{\cal O}(\frac{1}{s^2}). 
\end{equation}
 A solution regular at $s=0$ must fulfil
\begin{equation}
\label{Eor}
\frac{d\alpha}{ds} (0) = \frac{1}{3}(3/2-\epsilon+K e^{2\alpha(0)}) . 
\end{equation}
 These boundary conditions suggest a method of solution: For a given $K$ 
 we assume some $\epsilon$ and $C$ and, starting from the asymptotic values (\ref{Eas}) 
 at large $s$, integrate Eq.(\ref{Estat}) to $s=0$. 
 We check Eq.(\ref{Eor}) and the normalization and correct  
 $\epsilon$ and $C$ until we fulfil both.  
 
By using a new variable ${\bar v}=\frac{\partial f}{\partial Q}/(r \rho)$  
 and factoring out $2r\rho$, we transform Eq.(\ref{bou}) to a form 
\begin{equation}
\label{Edyn}
\frac{1}{2}{\ddot Q}{\bar v}+{\dot Q}^2(\frac{1}{2}\frac{\partial{\bar v}}{\partial Q}-
\frac{\partial\alpha}{\partial Q}{\bar v}+{\bar v}^2 [1+s(2\frac{\partial\alpha}{\partial s}-1)])=
\frac{\partial R}{\partial s} ,
\end{equation} 
 suitable for both instantons and oscillations if ${\dot Q}^2=2(V(Q)-{\cal E}_n)/B(Q)$ is understood.
$R[\alpha]$ stands for the l.h.s. of Eq.(\ref{Estat}). Let us call the l.h.s. of Eq.(\ref{Edyn}) 
$F$. If $F=0$ (no time dependence) we recover stationary solutions of Eq.(\ref{Estat}). 
For instantons, we solve Eq.(\ref{Edyn}) iteratively. Having a set of $\alpha(s,Q_i)$ 
 we calculate $F_i=F(s,Q_i)$. For each $F_i$ we  
  solve (\ref{Edyn}) as the ordinary differential equation in $s$ to obtain new $\alpha(s,Q_i)$. 
The method, as for the stationary case, is to adjust the asympotic form (\ref{Eas}) to the proper regularity 
condition at $s=0$. The new and old $F_i$-s are combined to provide initial $F_i$-s 
  for the next iteration.  
  With a careful modification of $F_i$-s this iteration leads to the self-consistency, i.e. 
 $F_i(old) = F_i(new)$. The initial densities $\rho_s(s,Q_i)$ are 
  obtained using the constrained imaginary-time step Hartree procedure. 
 We use the Runge-Kutta-Merson procedure for integration of Eqs.(\ref{Edyn}).
 Energies are calculated using a mesh of 128 equidistant points, $r/d_0=0-4.5$, and 
 the cubic spline interpolation for derivatives. We have checked that doubling the mesh density 
  does not change results in any appreciable manner.

\newpage 
 
Table 1 - Energies, crossover temperatures, decay exponents and rates 
for metastable and radially excited states of the $^7$Li BEC. Results 
 from bounce solutions ($^{*}$ from the functional minimization).  

\begin{center}

\begin{tabular}{c|cccccc} \hline

  $N$ & $\omega_n/\omega_0$ & $\frac{kT_c}{\hbar\omega_0}$ 
  & $\frac{S}{N}\times 10^3$ & $S$ & $\frac{\Gamma}{\omega_0}e^S$ & $\Gamma [s^{-1}]$ \\ \hline

 1255 &       & 0.130  & 2.6744 & 3.356 & 4.305  &  1.36 $\cdot 10^2$  \\ \hline

 1250 &       & 0.165  & 7.5520 & 9.440   & 4.756  &  3.43 $\cdot 10^{-1}$  \\
 n=1  & 0.967 & 0.188  & 2.5244 & 3.156   &        &  \\ \hline

 1245 &       & 0.189  & 13.097 & 16.306  & 6.87   &  5.17 $\cdot 10^{-4}$  \\
 n=1  & 1.086 & 0.208  & 7.6934 &  9.578  &        & \\
 n=2  & 2.112 & 0.217  & 3.5636 &  4.437  &        & \\ \hline

 1240 &       & 0.207  & 19.082 & 23.662  & 8.80  &  4.23 $\cdot 10^{-7}$ \\
 n=1  & 1.169 & 0.224  & 13.436 & 16.661  &        & \\
 n=2  & 2.292 & 0.234  & 9.0346 & 11.203  &        & \\
 n=3  & 3.373 & 0.241  & 5.1464 &  6.382  &        & \\ \hline

 1230 &       & 0.238  & 31.980 & 39.335  & 12.31  &  9.24 $\cdot 10^{-14}$  \\
 n=1  & 1.268 & 0.252  & 26.084 & 32.083  &        & \\
 n=2  & 2.518 & 0.260  & 21.347 & 26.257  &        & \\
 n=3  & 3.748 & 0.267  & 17.076 & 21.004  &        & \\ \hline

 1200$^{*}$ & & 0.305  & 75.40   & 90.490  & 21.12  &  9.63 $\cdot 10^{-36}$ \\ \hline 
 \end{tabular}

\end{center}

\vspace{2cm}

Figure captions

Fig.1  Potential energy $E(Q)=NV(Q)$ of BEC (in $\hbar\omega_0$) for various $N<N_c$.
       ($Q$ in units $d_0^2$)

Fig.2 Mass parameters $B(\sqrt{Q})=4QB(Q)$ from various instanton solutions,   
      overlayed in one picture. For gaussians, 4QB(Q)=1. 
   
Fig.3 
     Bounce penetrates the barrier practically in a finite $\tau$ (in units $\omega_0^{-1}$). 
  The metastable density $\rho(r,\pm\infty)$ is nearly equal to $\rho(r,\tau=3.43)$ shown. 

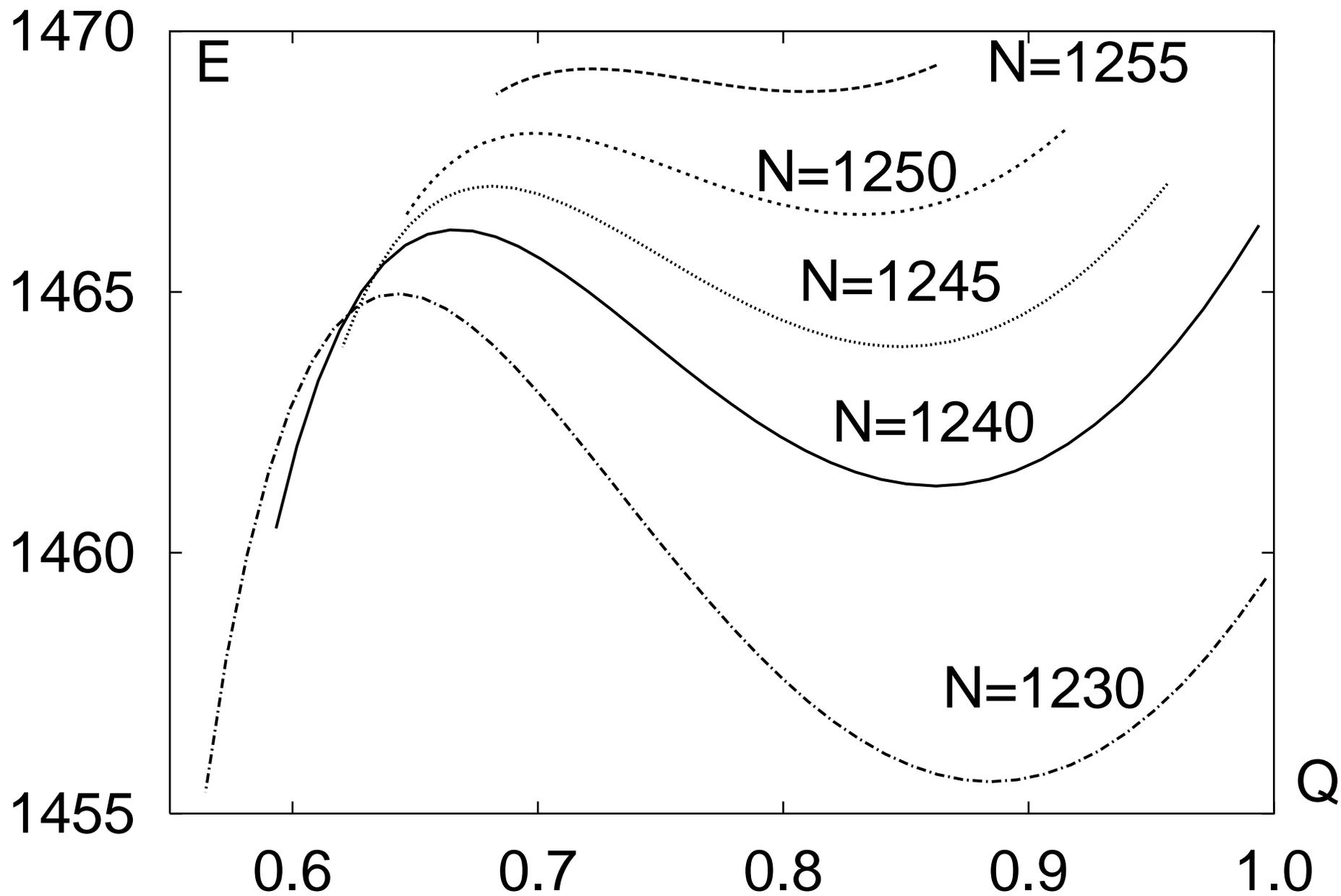
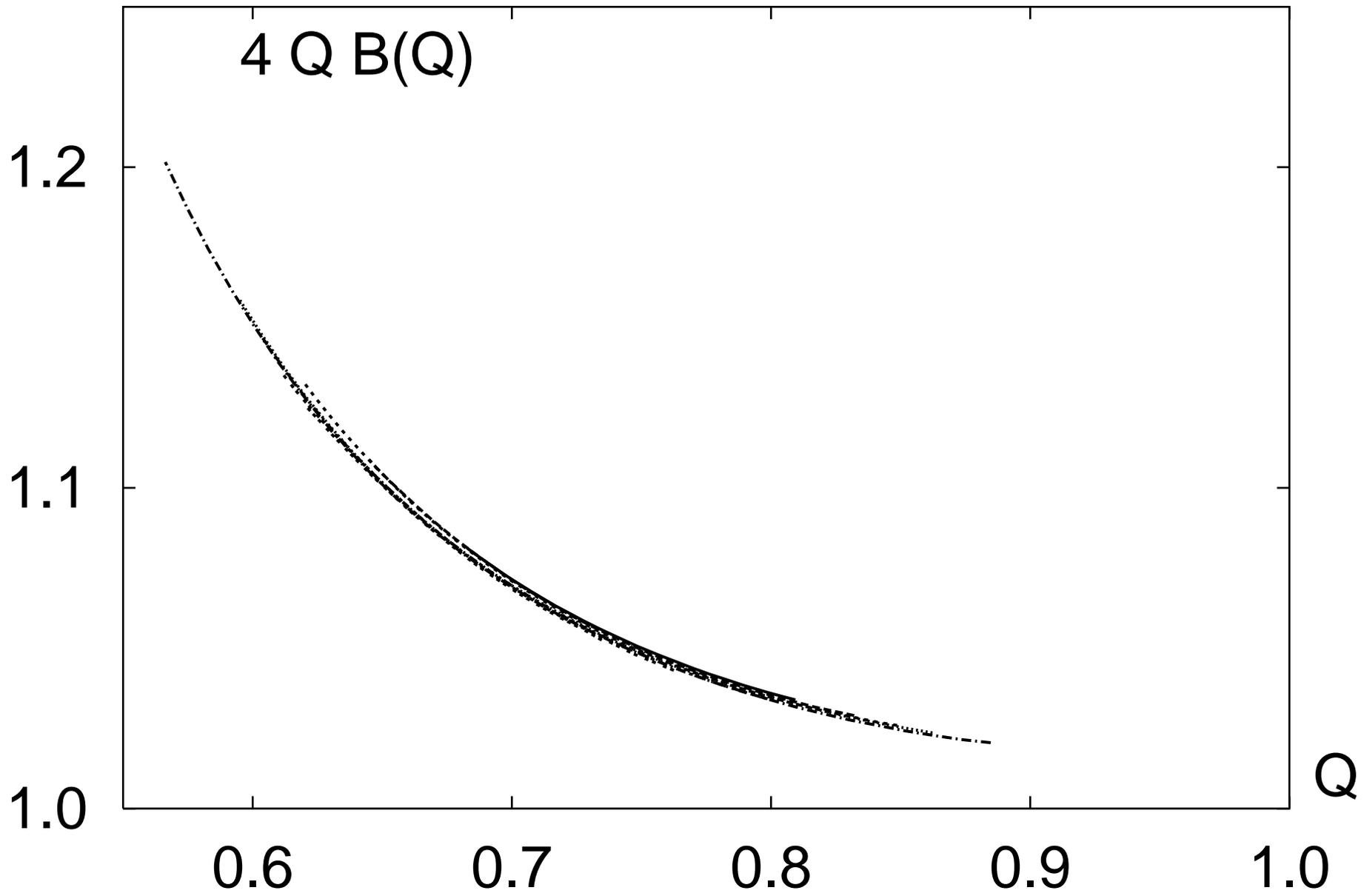
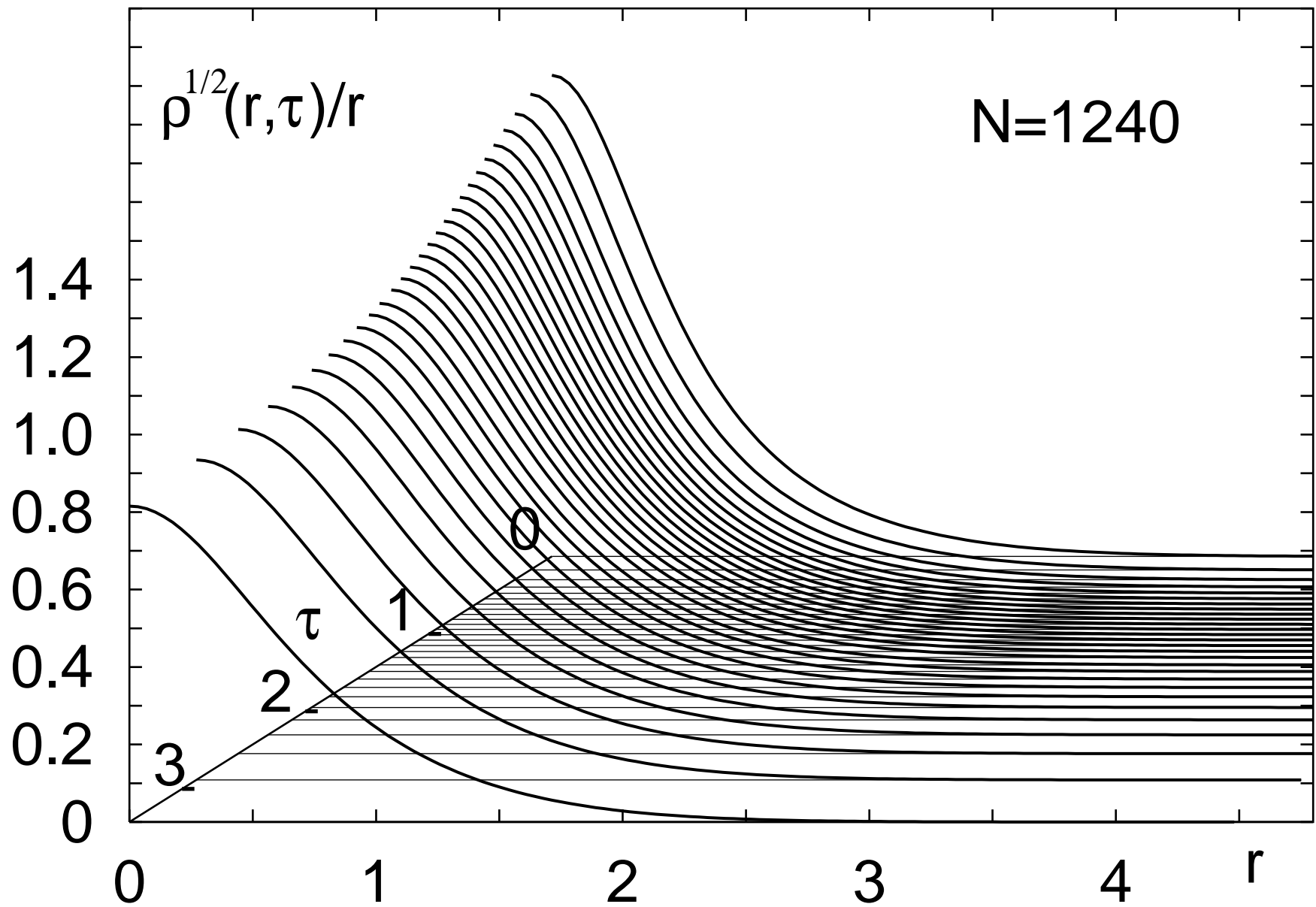
\end{document}